# Different Charges in the Same Bunch Train at the European XFEL

Yauhen Kot, Torsten Limberg and Igor Zagorodnov

*Deutsches Elektronen Synchrotron, Notkestrasse 85, 22607 Hamburg*

______________________________________________________________________

**Abstract**

The injector of the European XFEL was initially designed for the operation with 1nC bunch charges [1]. Later the flexibility of the nominal design of the injector with respect to the bunch charge was studied and extended also for smaller bunch charges down to 20pC [2]. A very tempting upgrade of this extension would be the operation of the European XFEL with different charges in the same train. It would make it suitable also for the experiments which require simultaneously different SASE pulse length or radiation power.

Operation of two bunches within the same train sets new requirements on the working points of the injector which are to be satisfied additionally to the ones of a single charge operation. From the beam dynamics point of view here is to mention the similarity of the beam optical functions after the first accelerating module and suitable for lasing shapes of both bunches in the train at the end of the linac. Due to different charges and thus to different space charge forces which act on bunches during the passage of the linac the last condition cannot be easily satisfied even if the similarity of optical functions at the beginning of the linac is achieved. A more subtle analysis of the interplay between mismatch of beam optical functions, emittance growth in the injector and different 6D beam dynamics in the linac is needed with the final goal of successful lasing of both charges.

In this paper we have investigated the possibility of the operation of different charges in the bunch train for the nominal design of the injector and for the case that it is extended by an additional laser system on the cathode. We have examined the problem of similarity of beam optical functions for different bunches in a train. We report also about the sensitivity of the beam optical functions on the chosen compression scenario and give an overview over the working points for the settings at the injector for single charge operation as well as combined working points for different bunch pairs.

______________________________________________________________________

**Introduction**

In this paper the operation of the European XFEL with different bunch charges within the same bunch train has been considered. In Section I the main aspects of the XFEL injector optics are discussed. In section II we give an overview of the working points for a single bunch operation for different bunch charges. In section III we report about the approach for the determination of the combined working point for two different charges. In section IV we consider the combined working points for different bunch charge pairs if an additional laser system for the generation of bunches at the cathode is available. Section V deals with the sensitivity of the combined working points on the phase of ACC1. Sections VI and VII show S2E and SASE simulations for the 250pC/500pC bunch pair for the nominal parameters of the XFEL injector.

**I. Optics Considerations**

*General remarks. Injector of the European XFEL*

From the beam optics point of view the XFEL injector may be divided into three parts. The first one begins at the cathode and is 14.48m long. There are no quadrupoles foreseen in this region. Manipulation of the beam optical functions is achieved by the choice of the solenoid and gun gradient settings as well as by means of RF focussing in the ACC1. Since the optics in this region is dominated by the space charge forces the same settings of the machine would deliver generally different beam optical functions for different charges.

In the second part of the injector the matching of the beam optics takes place. This matching section begins at the entrance into the first quadrupole at 14.48m and goes up to the end of the quadrupole I1.QI.4 which is placed at s=29.509m immediately after the laser heater. In this region the beam optical functions are subject to the initial settings of the gun (solenoid peak field, gun and ACC1 gradient and phase, rms laser beam size at the cathode, laser pulse length) though but they are considered already in the linear optics approximation and count as charge independent.

The third part of the injector begins after the matching point at s=29.509m where the optics is expected to be the same for all bunch charges and any initial settings of the gun.

*Dependence of the beam optics on laser beam size*

Under the assumptions for the optics at the XFEL injector mentioned above one has to guarantee that both charges within the same bunch train arrive the matching section with similar twiss functions. This is an important condition in order to avoid optics distortions like beta beat downstream of the matching point.

Figure 1 shows the simulations results for the beta function at the beginning of the matching section in dependence of the rms laser beam size on the cathode. The scan has been taken for the bunch charge of Q=20pC and the solenoid peak field of B=0.2220T. It represents the typical behaviour of the beta function. From the lower side the range of the allowed values for the laser beam size is determined by the 100% transmission point. This point goes prior to the local maximum of the beta function at XYrms=50μm.

Operation with smaller laser beam sizes leads to the deformation of the distribution in the phase space first which is later followed by the reduced transmission. The upper limit for the rms laser beam size is constrained by the conditions on the beam emittance which is required for the proper FEL operation [3] and suitable for the matching of the beam optical functions twiss parameters of the beam.

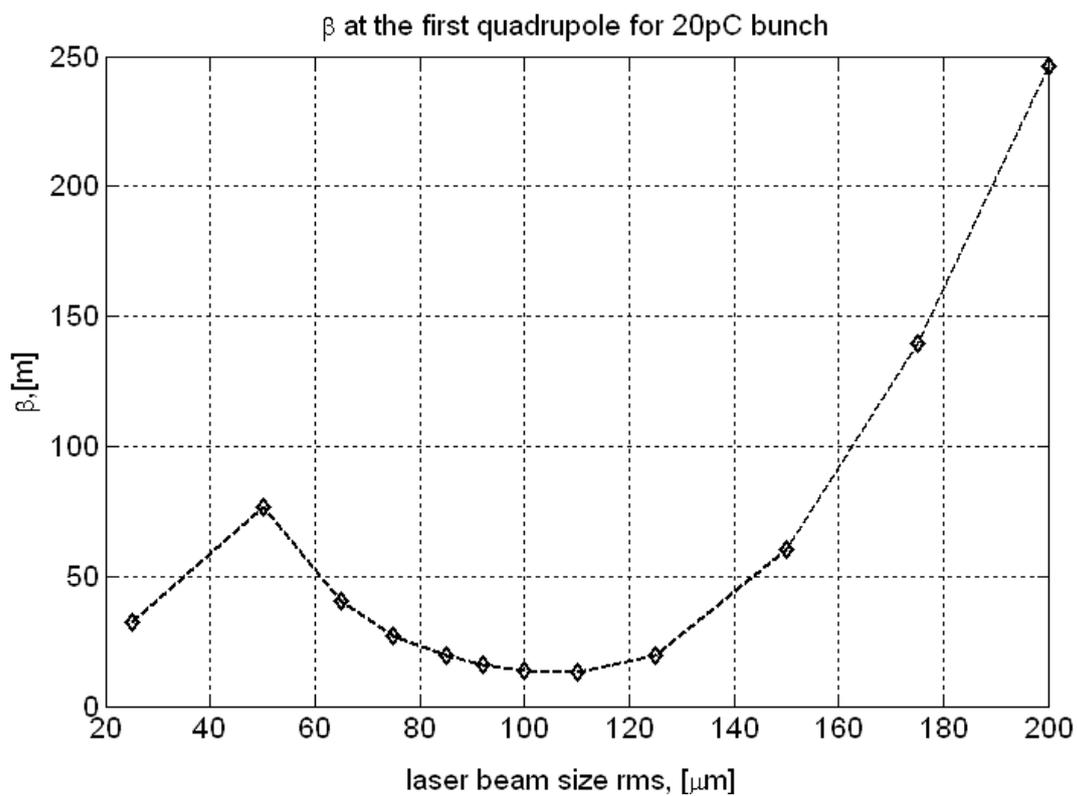

**Figure 1:** Dependence of the β-function at the beginning of the matching section on rms laser beam size. Simulations has been done for 20pC bunch charge and solenoid field of 0.2220T. Local maximum at 50μm indicates also the 100% transmission point. The emittance minimum has been found at 65μm.

Fig. 2 shows the scan of the beta function at the beginning of the matching section over the rms laser beam size for different bunch charges. The principal behaviour of the 20pC bunch from Fig. 1 is repeated also by higher charges while the whole curves are shifted to the right. This gives an opportunity to find gun settings which would provide the similar twiss parameters for two different bunch charges. In this case the choice of the laser beam size will be immediately after the local maximum on the descending branch of the curve for higher bunch charge.

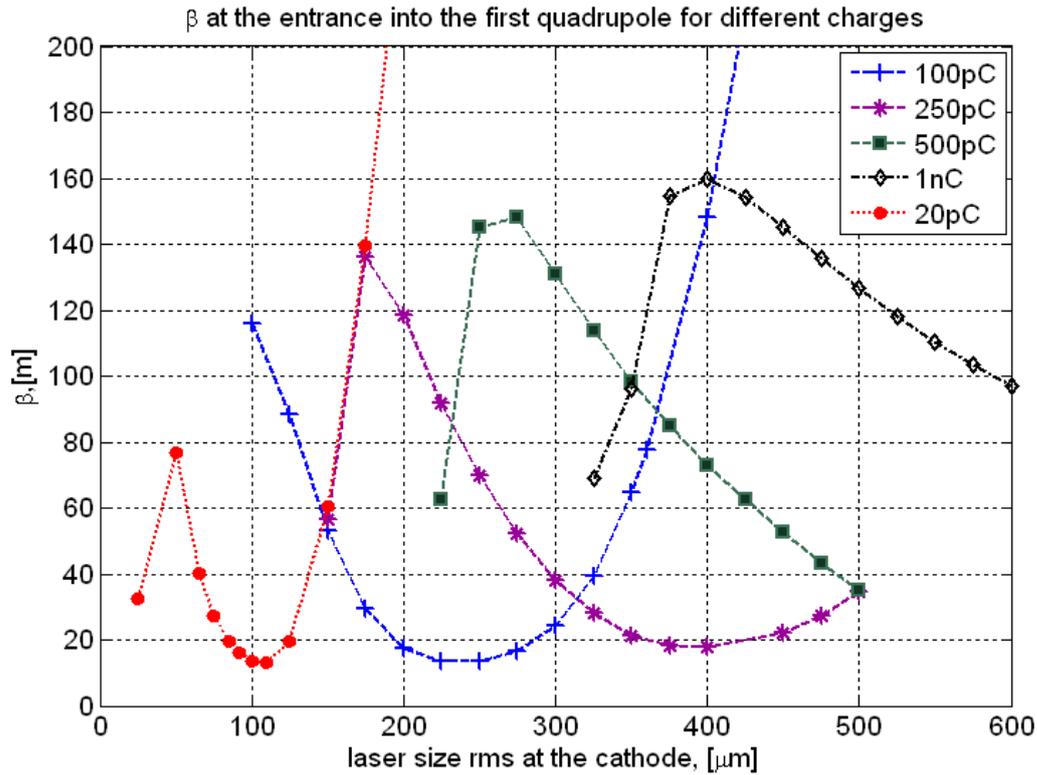

**Figure 2:** Dependence of the β-function on laser beam size rms for different bunch charges. Simulations are done for the peak solenoid field of 0.2220T. Combined working point for the operation with different bunch charges is imaginable for the laser sizes where the intersection of the descending arm of the curve of the higher charge and the ascending arm of the lower charge take place.

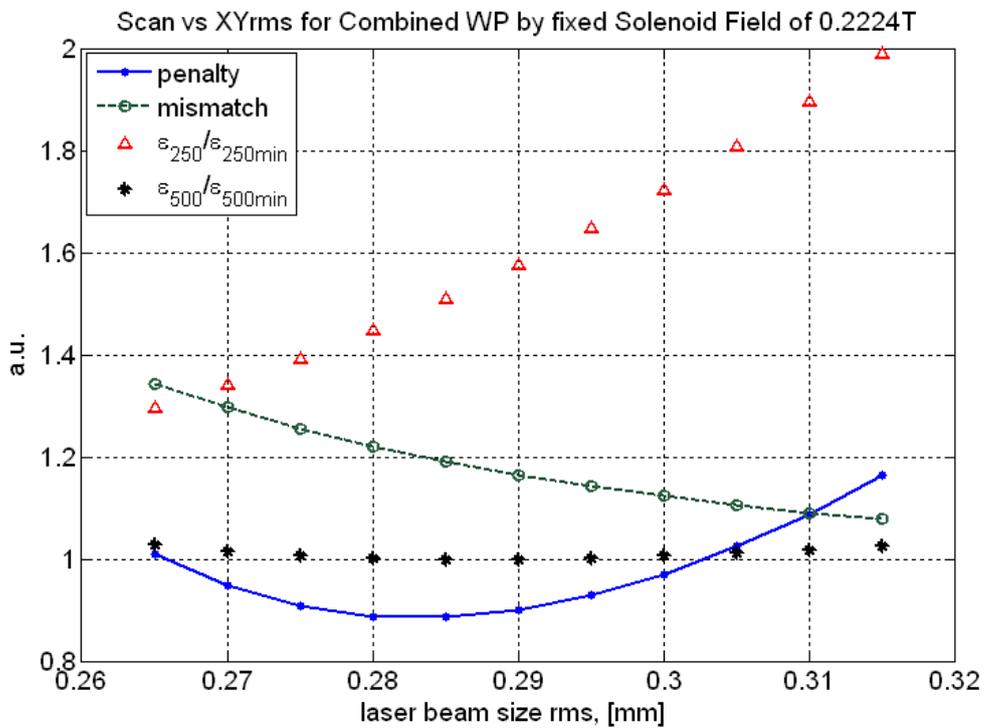

**Figure 3:** Scan over rms laser beam size for 250pC and 500pC bunch charges for fixed solenoid peak field of 0.2224T. Emittance development and the mismatch between beam optical functions of the both charges are shown. Combined working point was chosen according to the minimum of the penalty function (blue line) at XYrms=0.285mm.

This choice of the laser beam size leads to the emittance which is close to its minimum for the higher charge (Fig. 3). However it is already too large for the lower charge to keep the emittance comparably small as in the case of a single charge operation. Thus a combined working point for the setup of solenoid and laser beam size on the cathode is achievable for the operation of two different bunch charges on cost of the emittance growth of the lower bunch.

## II. Working Points for the Single Bunch Charge Operation

Determination of the working points for different bunch charges has been performed under the assumption of the same parameters of the gun and gun laser for all bunch charges. These parameters are summarized in Table 1. We have also fixed in the simulations the energy of the beam to 150MeV at the exit of the ACC1 section and to 130MeV after the ACC39. The energy gain pro cavity has been assumed to be equal for all cavities inside the ACC1 and ACC39. Since the compression scenarios require an off crest operation already at ACC1 the choice of the compression scenario will also have an impact on the beam optical functions at the entrance into the first quadrupole. Sensitivity of the beam optical functions to the phase of the ACC1 is discussed in the section V. In this section we assume the off crest operation of the ACC1 with the phase of 17.8 degrees and gradient of ACC1 adjusted in the order that the energy of the beam reaches 150MeV after ACC1.

| Table 1: Gun Parameters | | |
|---|---|---|
| Gun Gradient | Gun Phase | Laser Pulse Length and Form |
| 60MeV/m | -1.9 | Flat top 20ps, rise and fall time 2ps |

A two dimensional scan over solenoid peak field and laser beam size rms on the cathode has been undertaken in order to find the suitable working point for the operation of the XFEL injector. The simulation for each point has been done by means of the ASTRA tracking code using rotational symmetric algorithm with 200000 particles. The mesh has been chosen to be NradxNlong=40x100. This choice of numeric parameters has been recommended in [4] as a lower limit at which numeric effects do not play a significant role anymore.

| Table 2 : working points and summary over beam parameters at the beginning of the matching section for the operation of the XFEL injector with different bunch charges | | | | | |
|---|---|---|---|---|---|
| | Working Point | | Emittance and twiss functions at $1^{st}$ quadrupole (s=14.44m from cathode) | | |
| Q, [pC] | MaxB1, [T] | XYrms, [mm] | $\varepsilon$, [m$^{-6}$] | $\beta$, [m] | $\alpha$ |
| 20 | 0.2196 | 0.057 | 0.0824 | 132.00 | -9.483 |
| 100 | 0.2206 | 0.100 | 0.1925 | 161.10 | -11.21 |
| 250 | 0.2220 | 0.180 | 0.2982 | 42.13 | -2.305 |
| 500 | 0.2224 | 0.285 | 0.4391 | 19.81 | -0.6045 |
| 1000 | 0.2226 | 0.440 | 0.7091 | 8.729 | 0.0299 |

Table 2 summarizes the working points for the examined bunch charges which turned out to provide the smallest emittance in each particular case. Since the problem description remains up to the first quadrupole rotationally symmetric the values of the emittance and beam optical functions are presented only for the horizontal plane.

Although the working points from Table 2 provide the best emittance they are not necessarily suitable for the operation of the XFEL injector. The ones for low bunch charges like 20pC and 100pC seem to be for instance not practical because of too large values of the beam optical functions at the exit of the first accelerating module. In these cases one can get no reasonable matching of the beam optics to the FODO diagnostics section at s=29.5m. The problem of the matching ability has been investigated by means of the MAD8 matching algorithm. We found out that the condition of $\beta$<100m and $|\alpha|$<5 immediately after the first accelerating module at s=14.44m is a good guess for the matching ability of the beam optical functions. In the Fig. 4 the operation windows for the settings of the solenoid peak field and the rms laser beam size are shown for the charges of 20pC and 100pC. The choice of the working point from the proposed window allows the match of the beam optical functions on one hand and keeps the increase of the emittance below 10% compared to its minimum on the other hand.

Higher bunch charges like 250pC, 500pC and 1nC can be operated at the working points from the Table 2. For these charges we have found that the settings of the solenoid peak field and the rms laser beam size which lead to the smallest possible emittance provide also suitable conditions for the matching of the beam optical functions.

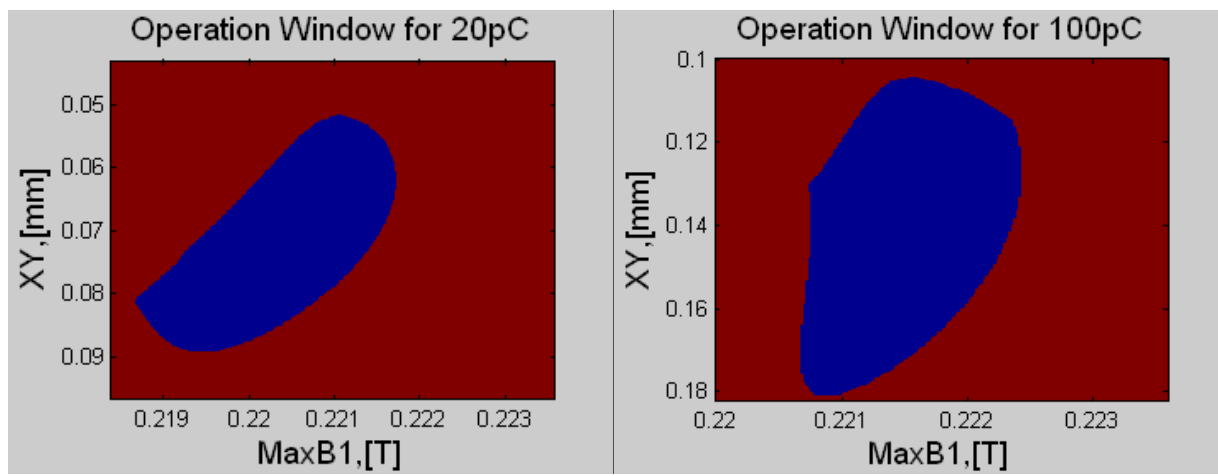

**Figure 4: Operation window for 20pC and 100pC bunch charges. For the working points from the blue region the conditions of matching ability of the beam optical functions and emittance growth of less than 10% compared to the minimum hold. Working points which provide the minimum of the emittance lie outside the blue region for both charges.**

## III. Determination of the Combined Working Points for Two Different Bunch Charges in the Same Bunch Train

We have assumed the same operation conditions as in the previous chapter for the determination of the combined working points. Under combined working points we understand here the settings of the solenoid peak field and rms laser beam size on the cathode which act on both bunches in the bunch train. In this section we consider cases in which neither solenoid peak field nor laser beam size cannot be tuned individually for any particular bunch charge.

In order to find the combined working points we have analyzed the intersections of the operation windows for different bunch charges which have been considered in the previous section. Except of the pair of 250pC/500pC a combined operation of two different bunch charges with the same solenoid field and rms laser beam size appeared possible only for the pairs of 100/250pC and 500pC/1nC. There is no common operation window available for other pairs of bunch charges.

As it was already mentioned in the Section I the condition of the similarity of the beam optical functions can be fulfilled on cost of the large emittance growth of the lower charge. For example, the comparison of the scans for 250pC and 500pC charges has shown a perfect agreement of beam optical functions for MaxB=0.2235T and XY=0.325mm. However it is achieved on cost of emittance growth for the 250pC bunch of about 160% compared to the possible minimum. Recent tests at FLASH [5] related to the operation of the two bunch charges in the same train has shown however that the distortion of the beam optical functions which has been introduced by the change of the energy gain by 5% and keeping at the same time the currents in the quadrupole magnets unchanged leads though to a SASE drop of 10-20% but operation and generation of SASE remain acceptable. This makes reasonable to take also the emittance growth into account while looking for the optimal combined working point.

We introduce the penalty function of the form

$$pen = \frac{\Delta \epsilon_1}{\epsilon_{10}} + \frac{\Delta \epsilon_2}{\epsilon_{20}} + 2\ \xi_{1\to 2} - 1$$

where $\epsilon_1, \epsilon_2$ emittance of both bunches, $\epsilon_{10}, \epsilon_{20}$ found minima of the emittance and $\xi_{1\to 2} = 0.5\ \beta_1\gamma_2 - 2\alpha_1\alpha_2 + \beta_2\gamma_1$ the mismatch parameter between the beam optical functions of both bunches evaluated at the entrance into the first quadrupole and define the combined working point as the pair of values (MaxB, XYrms) with MaxB the solenoid peak field and XYrms the rms laser beam size on the cathode which minimizes this penalty function.

The results of the scan comparison for bunch pairs mentioned above are shown in Figures 5-6 and summarized in Table 3.

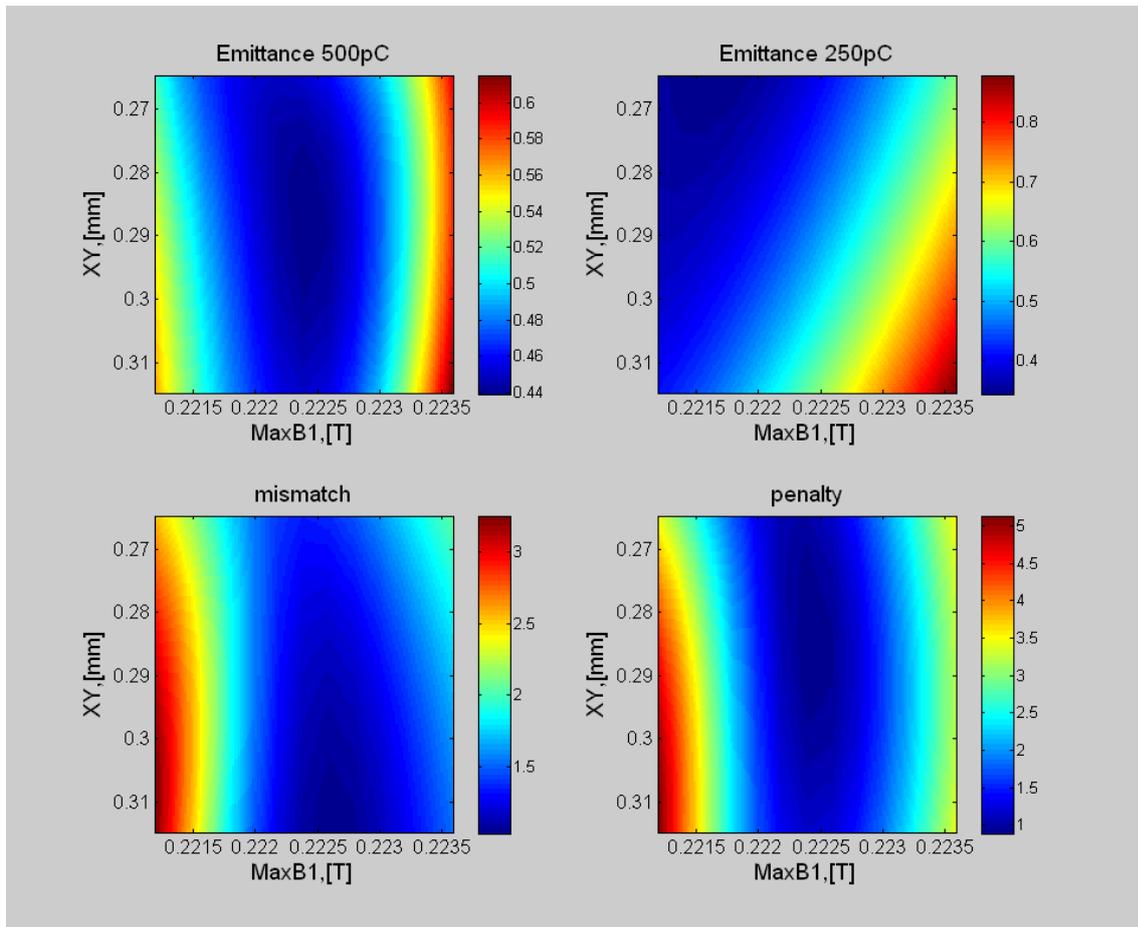

**Figure 5: Scan over working points for the combined operation of 500pC and 250pC bunch charges in the nominal design of the XFEL injector**

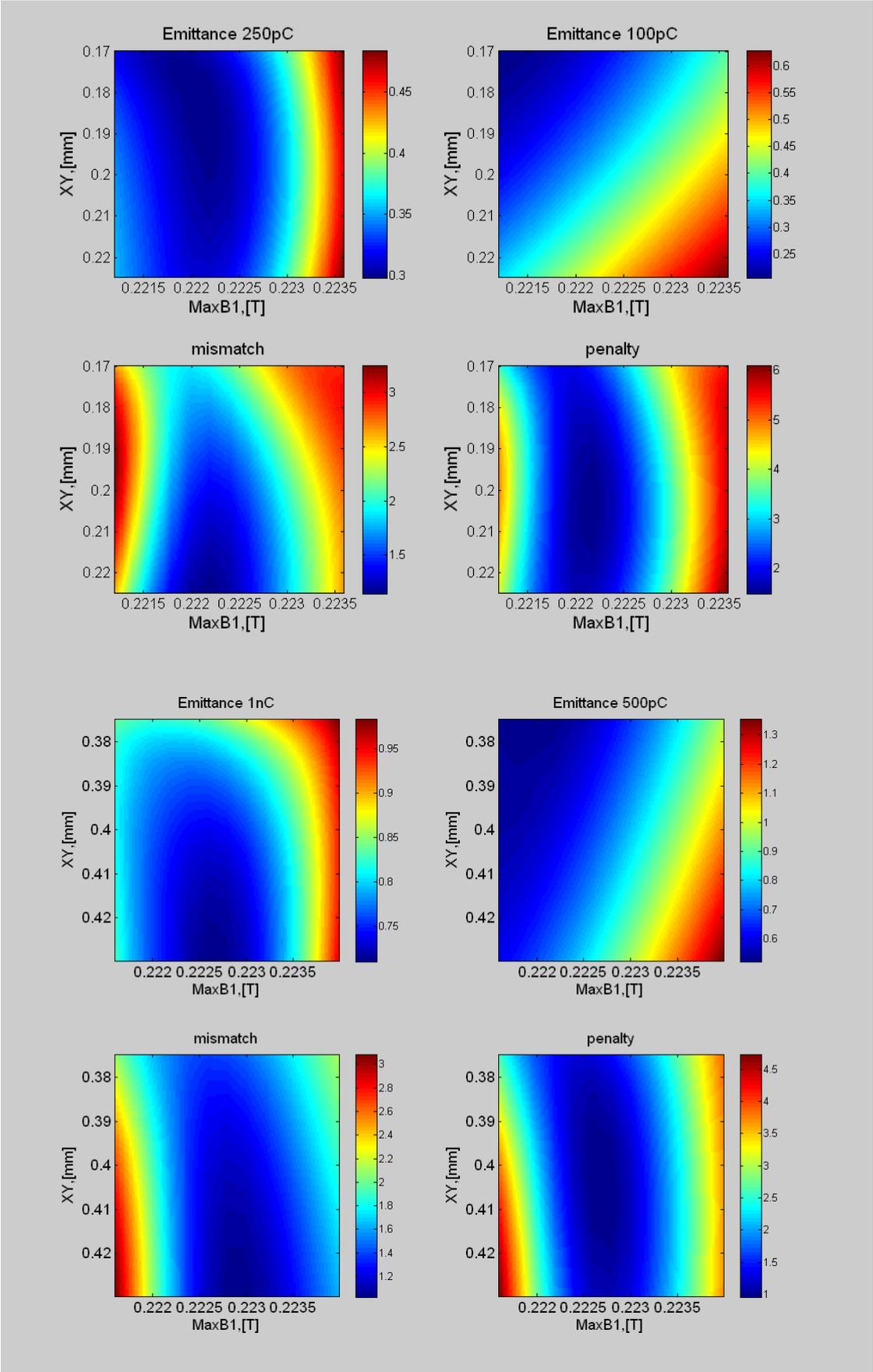

**Figure 6: Combined working points for 250pC/100pC bunch pair (upper four pictures) and for 1nC/500pC bunch pair (lower four pictures)**

| Table 3: Combined working points for different bunch charge pairs. Basic design of the XFEL injector | | | | | | |
|---|---|---|---|---|---|---|
| $Q_1/Q_2$, [pC] | WP | | $\Delta\varepsilon_1$, % | $\Delta\varepsilon_1$, % | $\xi_{1\to 2}$ | pen |
| | MaxB1, [T] | XY, [mm] | | | | |
| 100/250 | 0.2220 | 0.2050 | 74.8 | 1.7 | 1.403 | 1.4871 |
| 250/500 | 0.2224 | 0.2850 | 50.1 | 0.0 | 1.192 | 0.8869 |
| 500/1000 | 0.2226 | 0.4000 | 51.7 | 4.1 | 1.194 | 0.9569 |

**IV Combined Working Points for the Operation with Two Laser Systems**

An additional laser system means for the simulations that both bunch charges in the train experience the same solenoid field though however the laser beam size on the cathode can be adjusted for each charge individually. In this case it becomes possible to operate any bunch pair within the same bunch train.

In Figures 7 - 8 the scan results for the bunch pairs from the previous chapter are shown. For each value of the solenoid peak field we have looked for the minimum of the penalty function. The possibility to adjust the laser beam size on the cathode for each bunch charge separately leads to a significant improvement of the bunch parameters. Thus the penalty function for the bunch pair of 500pC/250pC can be reduced from 0.8869 to 0.1058. The emittance growth remains below 3% for each charge and the mismatch of the beam optical functions is reduced from 1.192 to 1.026. Working points and results for this case as well as for other bunch pairs are summarized in Table 4.

| Table 4: Combined working points for XFEL injector with two laser systems on the cathode | | | | | | | |
|---|---|---|---|---|---|---|---|
| $Q_1/Q_2$, [pC] | MaxB,[T] | $XY_1$,[μm] | $XY_2$,[μm] | $\varepsilon_1$,[m$^{-6}$] | $\varepsilon_2$,[m$^{-6}$] | Mism. $\xi$ | pen |
| 1000/500 | 0.2226 | 395 | 310 | 0.7492 | 0.4506 | 1.0675 | 0.1502 |
| 1000/250 | 0.2228 | 380 | 240 | 0.8156 | 0.3765 | 1.0902 | 0.5859 |
| 1000/100 | 0.2230 | 375 | 167.5 | 0.8714 | 0.3167 | 1.0647 | 0.9916 |
| 1000/20 | 0.2230 | 390 | 81 | 0.7854 | 0.1384 | 1.0448 | 0.8139 |
| 500/250 | 0.2224 | 265 | 210 | 0.4522 | 0.3069 | 1.0264 | 0.1058 |
| 500/100 | 0.2226 | 265 | 152.5 | 0.4596 | 0.2430 | 1.0451 | 0.3898 |
| 500/20 | 0.2228 | 270 | 71 | 0.4648 | 0.1229 | 1.0270 | 0.5880 |
| 250/100 | 0.2220 | 170 | 122.5 | 0.3031 | 0.1949 | 1.0342 | 0.0842 |
| 250/20 | 0.2226 | 200 | 69 | 0.3137 | 0.1158 | 1.0412 | 0.5187 |
| 100/20 | 0.2222 | 137.5 | 67 | 0.2019 | 0.1032 | 1.0612 | 0.4025 |

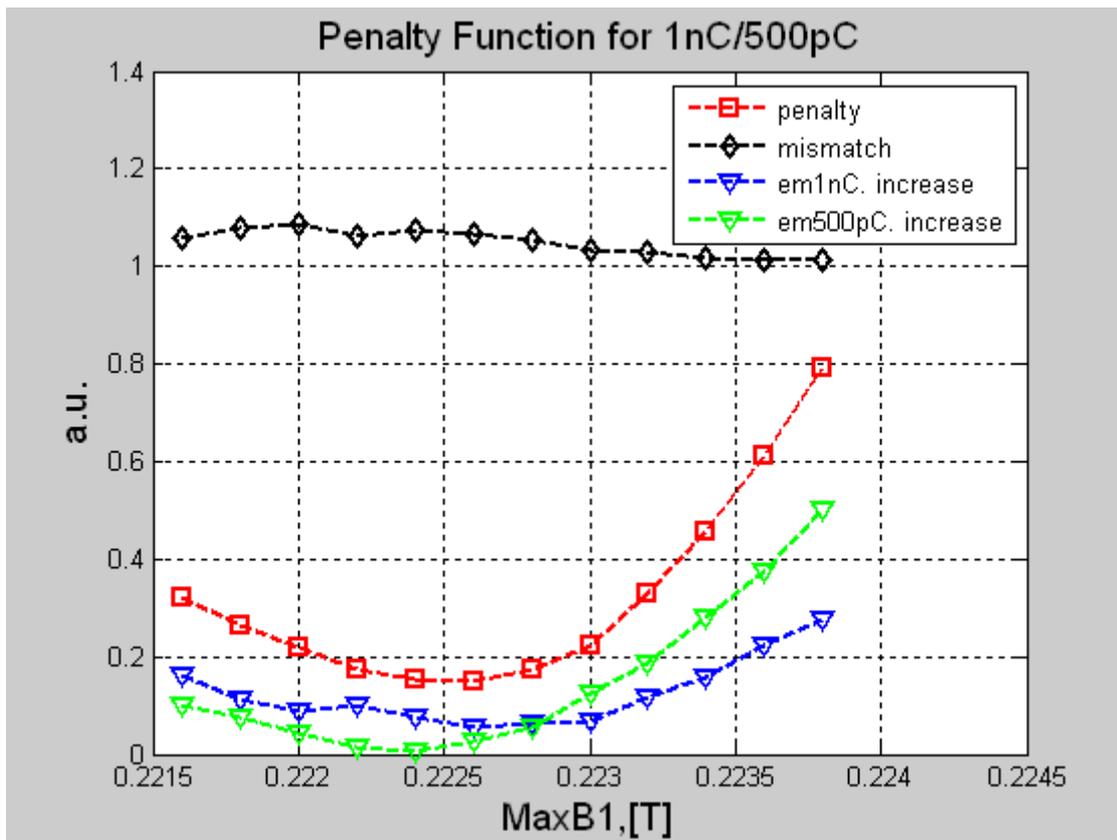

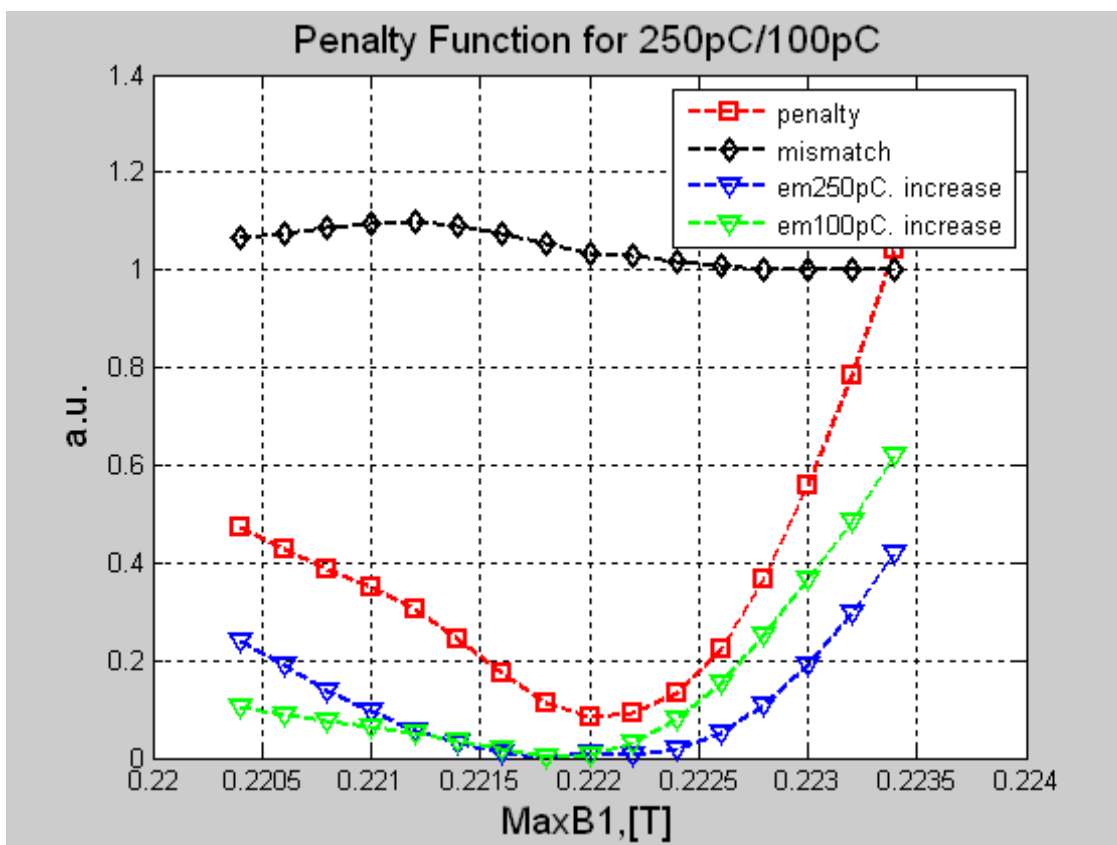

**Figure 7: penalty function and other parameters for 1nC/500pC bunch charge pair (upper plot) and 250pC/100pC pair (lower plot) if two laser systems are available at the cathode.**

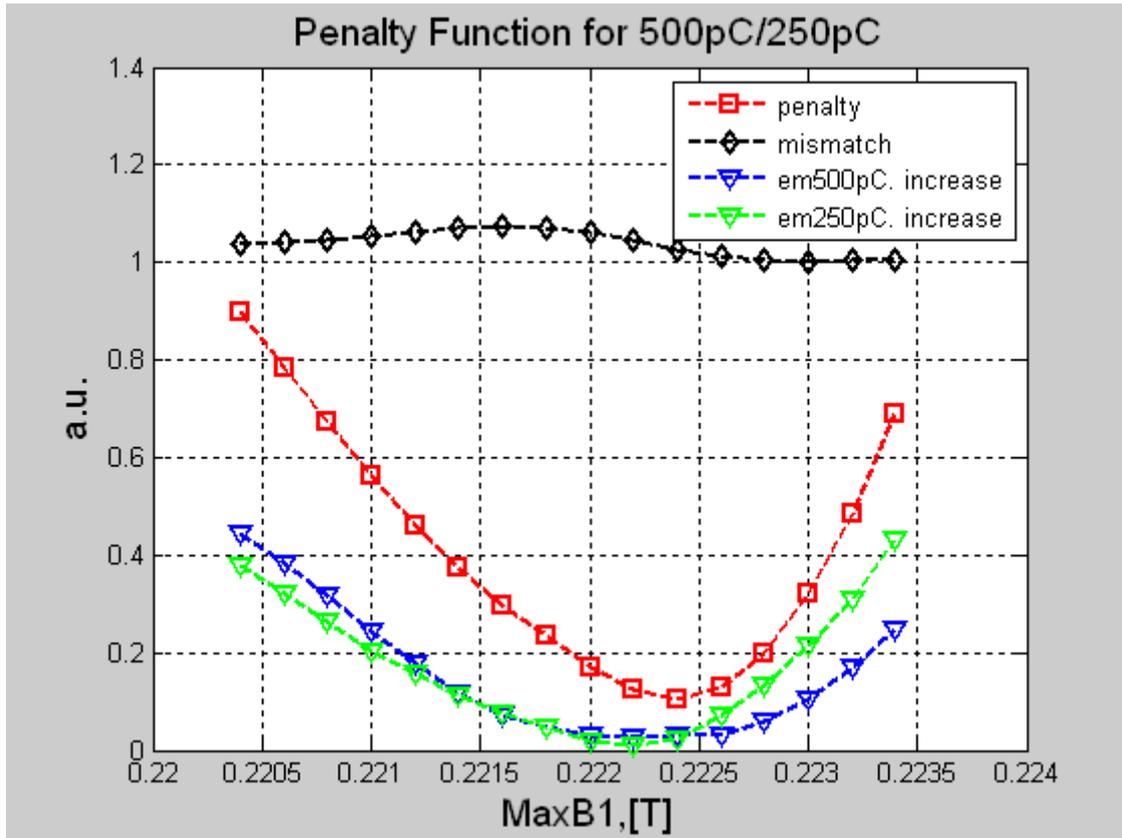

**Figure 8:** penalty function and other parameters for 500pC/250pC bunch charge pair if the second laser system is available at the cathode. Minimum of the penalty function is found for MaxB1=0.2224T leading to the emittance of 0.4522m$^{-6}$ and 0.3069m$^{-6}$ correspondently. Compared to the minimum the emittance increase remains below 3% for both charges. The mismatch of the beam optical functions between the charges is 1.0264.

### V. Sensitivity of the combined working points to RF settings

Since the compression takes place already in the ACC1 upstream to the matching section, a complementary 2D scan over the gradient and the phase of the ACC1 has been performed in order to verify the sensitivity of the beam optical functions with respect to compression scenarios. The results of this scan are shown on Figure 9. With the penalty function introduced above they determine the operational window of the RF settings for the possible compression scenarios approximately as [125:160] MeV for the maximum energy gain and [0:18] degrees for the possible off crest operation. Together with the values for solenoid field and rms laser beam size from Table 3 this defines the combined working point for the simultaneous operation of 250pC and 500pC bunches. The ACC1 settings for the on crest and off crest operation which has been used in the simulations are summarized in Table 5.

| Table 5: ACC1 Settings assumed in the simulations | | | |
|---|---|---|---|
| On-crest settings | | Off-crest settings | |
| $\Delta E_{max}$,[MeV] | Phase,[deg] | $\Delta E_{max}$,[MeV] | Phase,[deg] |
| 145,04 | 0,0 | 157,29 | 17,8 |

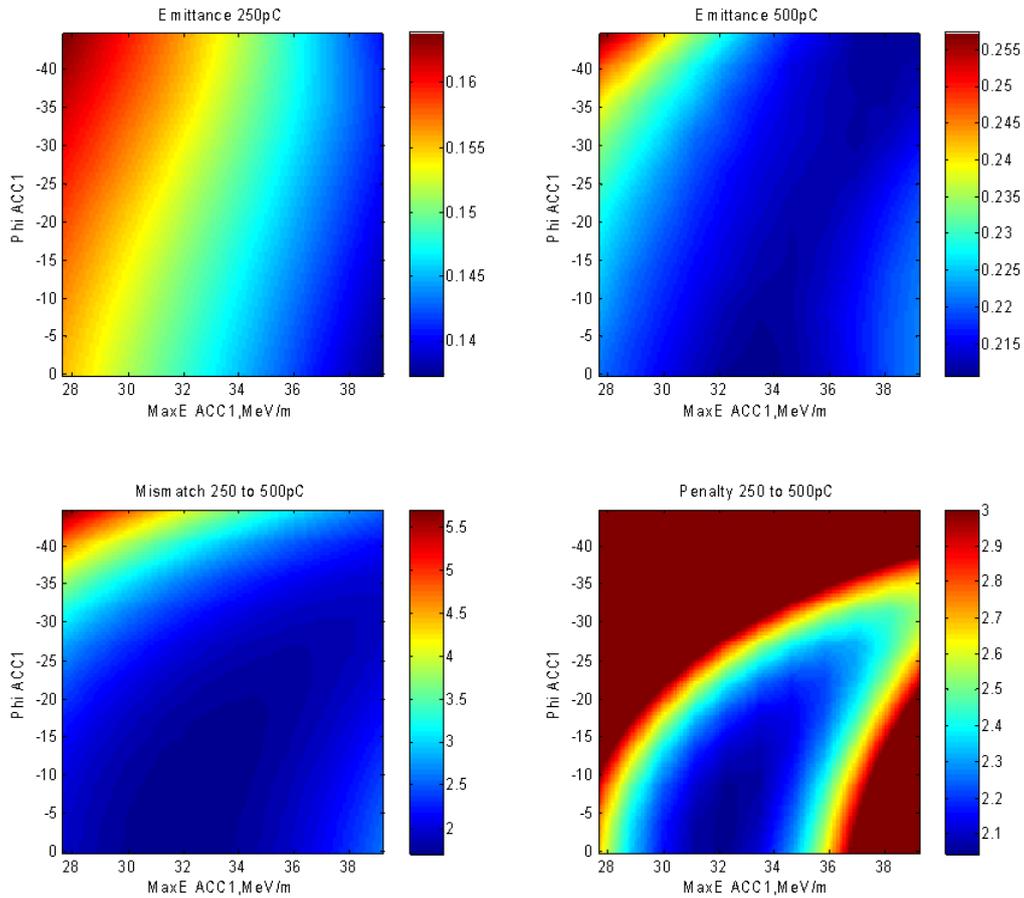

**Figure 9:** 2D Scan over gradient and phase of the ACC1 for fixed values of the solenoid field and laser beam size (determined combined working point for the nominal design of XFEL injector). Last picture shows the final operation window.

**VI Operation with 250pC and 500pC bunch charges within the same train. RF Adjustment**

We consider an operation of 500pC and 250pC bunches within the same bunch train for the nominal design of the XFEL injector i.e. with the same solenoid field and rms laser beam size for both charges.

| Table 6: input parameters for the compression scheme | | | | | | | | |
|---|---|---|---|---|---|---|---|---|
| Compression factors | | | R56 at bunch compressors, [m] | | | Energy Profile, [MeV] | | |
| C1 | C2 | C | R56_0 | R56_1 | R56_2 | E0 | E1 | E2 |
| 3 | 25 | 300 | 0.065 | 0.043 | 0.020 | 130 | 650 | 2300 |

The first rough definition of the RF settings has been accomplished by means of RF tweak tool [6]. The tool simulates the dynamics of the particle distribution mainly in the longitudinal phase space, takes into account some collective effects like the longitudinal space charge and CSR and gives also a simple approximation of transverse dynamics. It is based on the semi analytical approach which has been described in [7]. This way it becomes possible to find the first guess of the proper rf settings which would provide the chosen compression scheme.

For the input parameters of the compression scheme which are summarized in the Table 6 we have tried to establish a suitable Gaussian-like profile for both bunch charges with a peak current of several kA at the end of the Linac 3 (at s=1630m from the cathode).

Once the "first guess" for the RF settings was found the expected reference distributions after each bunch compression stage have been saved and a start to end run has been performed for both bunch charges from the first quadrupole up to the end of the Linac 3. The longitudinal particle distributions from the start to end run normally differ from the reference distributions due to more accurate calculation of the collective effects and taking into account of the wake fields. In order to get the bunch shape in the longitudinal phase space close enough to the reference distribution an additional RF adjustment was performed. Table 7 shows the final RF settings which provide reasonable final longitudinal distributions with a peak current of 7.75kA for 250pC and 5.34kA for 500pC bunch charge.

The main beam parameters at the beginning and at the end of the S2E simulations are summarized in Tables 8a and 8b. The detailed forms of the longitudinal profiles for both bunches are shown in Fig. 10 and Fig. 11.

| Table 7: RF settings for the simultaneous compression of 250pC and 500pC bunch charges | | | | | |
|---|---|---|---|---|---|
| | ACC1 | ACC39 | L1 | L2 | L3 |
| $\Delta E_{max}$, [MeV] | 157.29 | 26.03 | 657.49 | 1667.35 | 11764,04 |
| Phase,[deg] | 17.79 | 186.13 | 37.69 | 6.57 | 0.00 |

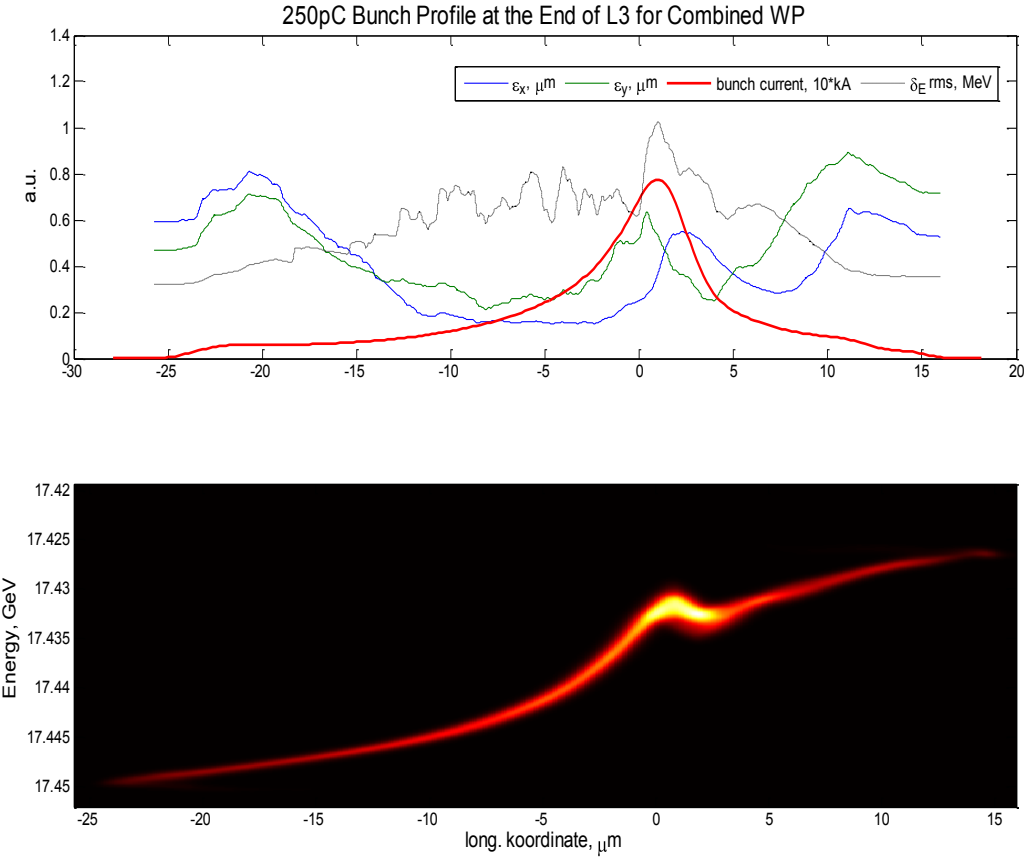

**Figure 10:** Summary about the 250pC bunch after the passage through the S2E section at the combined with 500pC bunch working point

| Table 8a: some beam parameters for the operation at the combined working point evaluated at the beginning (start) and end of the S2E section |||||||||
|---|---|---|---|---|---|---|---|---|
| | 250pC |||| 500pC ||||
| | $\varepsilon_{x,pr}$, [μm] | $\varepsilon_{y,pr}$, [μm] | $\varepsilon_{x,core}$, [μm] | $\varepsilon_{y,core}$, [μm] | $\varepsilon_{x,pr}$, [μm] | $\varepsilon_{y,pr}$, [μm] | $\varepsilon_{x,core}$, [μm] | $\varepsilon_{y,core}$, [μm] |
| Start | 0.336 | 0.325 | 0.168 | 0.168 | 0.424 | 0.422 | 0.394 | 0.394 |
| End | 0.391 | 0.509 | 0.311 | 0.404 | 0.486 | 1.367 | 0.414 | 0.550 |

| Table 8b: some beam parameters for the operation at the combined working point evaluated at the beginning (start) and end of the S2E section |||||||||
|---|---|---|---|---|---|---|---|---|
| | 250pC |||| 500pC ||||
| | $\delta_{E,rms}$ [MeV] | $I_p$, [A] | $\tau$, [fs] FWHM | C | $\delta_{E,rms}$ [MeV] | $I_p$, [A] | $\tau$, [fs] FWHM | C |
| Start | 2.47e-4 | 16.12 | 11225 | 196 | 7.58e-4 | 26.14 | 13890 | 136 |
| End | 0.765 | 7.75e+3 | 19 | | 0.553 | 5.34e+3 | 78 | |

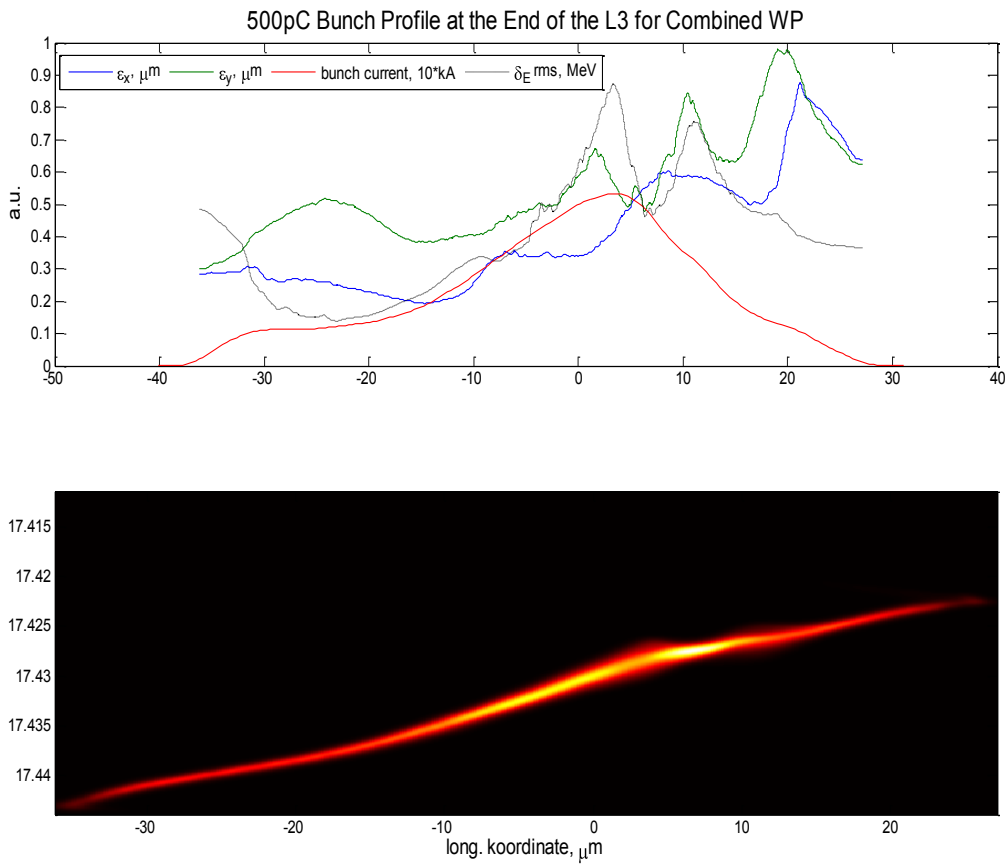

Figure 11: Summary about the 500pC bunch after the passage through the S2E section at the combined with 250pC bunch working point

## VII SASE Simulations for the bunch charge pair 500pC/250pC

Lasing simulations for the SASE 1 section has been performed by means of the Genesis tracking code [8]. The beam input files have been generated from the output files of the S2E simulations while the evolution of the beam optical functions between Linac 3 and undulator section has been reproduced by means of the linear optics approximation using the XFEL lattice file.

The simulations indicate the resulting SASE pulse energy of 618µJ for 250pC and 708µJ for 500pC bunch charge. Though both bunch charges demonstrate comparably equal pulse energy in the same time they differ in pulse length and average laser power. The part of the bunch which contributes to the SASE process is three times longer for 500pC bunch than that for 250pC bunch (approximately 30µm compared to 10µm). On the other side the 500pC bunch has smaller peak current (5.34kA compared to 7.58kA) and also significantly smaller maximum averaged power (12.5GeV compared to 31GeV for 250pC bunch charge). Moreover the laser pulse shape shows a double horn structure with a high peak power in case of 250pC bunch charge while the 500pC bunch keeps a stable "flat top" pulse with moderate laser power over the entire length of 30$\mu$m (see Fig. 10 and 11).

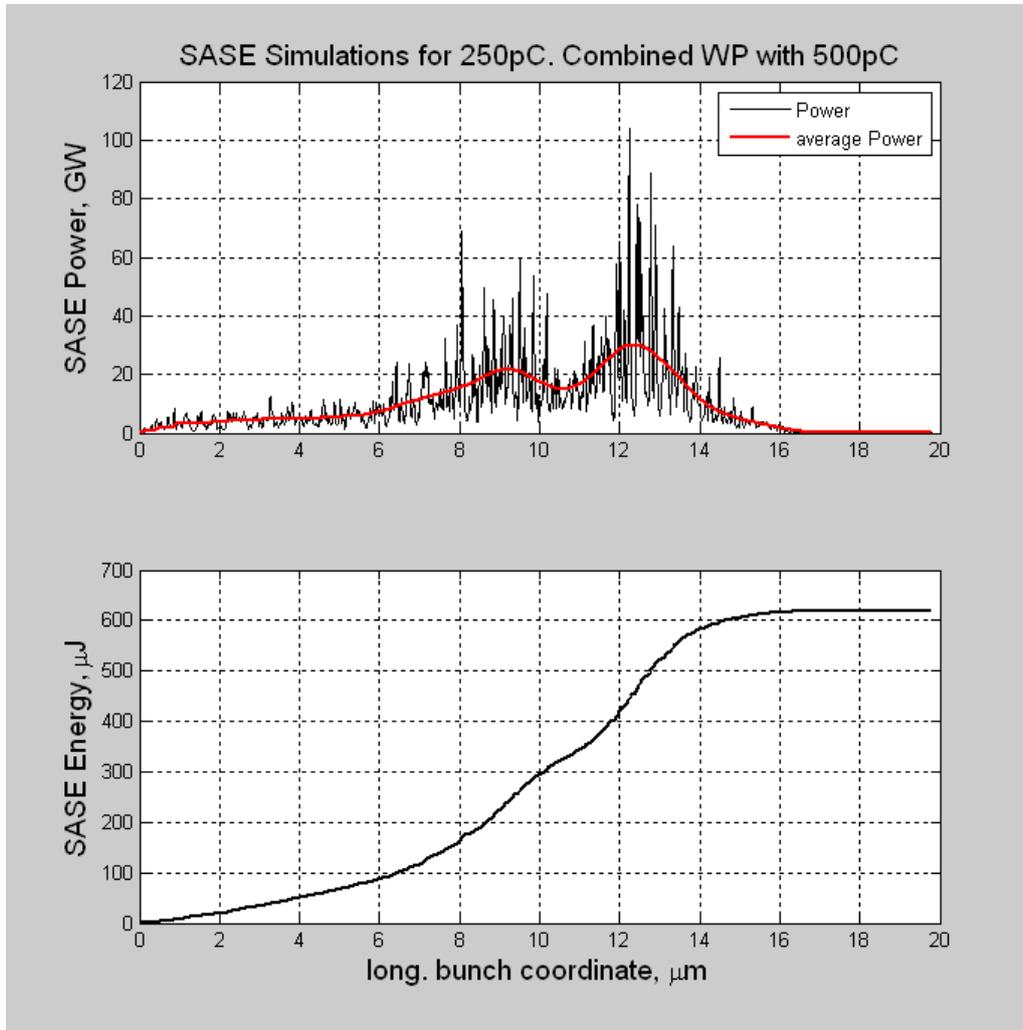

**Figure 10: SASE Simulations for 250pC bunch operated at the combined working point.**

| Table 9: Results of SASE simulations for the operation of 500pC and 250pC bunch charges in the same bunch train for the nominal design of the XFEL Injector ||||| 
|---|---|---|---|---|
| Bunch charge,[pC] | $I_{Peak}$,[kA] | Pulse Energy [$\mu$J] | Max($P_{av}$),[GW] | Pulse Length $\tau$,[$\mu$m] |
| 250 | 7.58 | 618 | 31.0 | 10 |
| 500 | 5.34 | 708 | 12.5 | 30 |

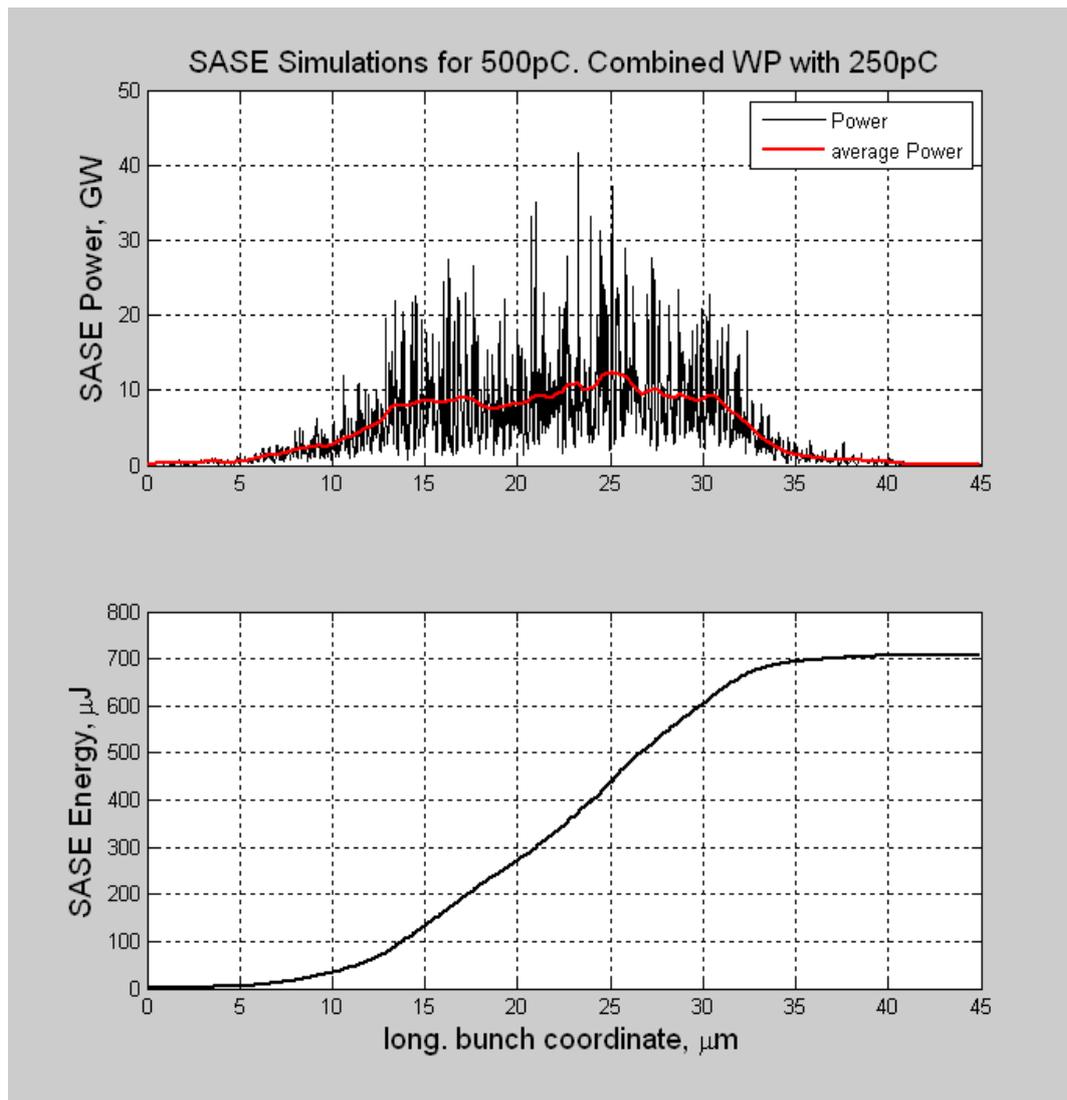

Figure 11: SASE Simulations for 500pC bunch operated at the combined working point

**Summary**

Crucial point for the operation of the European XFEL with two different bunch charges within the same bunch train is the ability to achieve similar beam optical functions of both bunches after the first accelerating module. Simulations have shown that for the nominal design of the XFEL injector (both bunch charges experience the same solenoid field and laser beam size at the cathode) it is possible for some bunch pairs on cost of the emittance growth of the bunch with the smaller charge. In order to get suitable bunch parameters it is reasonable however to look for a compromise between emittance growth and mismatch of the beam optical functions between bunches. In this paper we have proposed a penalty function which takes into account these both effects. Combined working points have been defined and derived by means of the simulations with ASTRA code as ones which minimize the penalty function.

A complete S2E simulations has been performed for the combined working point for the operation of the 250pC and 500pC bunch charges. It was found that for the nominal design of the XFEL i.e. under the same settings of the solenoid peak field, rms laser beam size at the cathode and the same rf settings of the acceleration sections it is possible to achieve bunch shapes which are suitable for the lasing for both bunch charges. SASE simulations have been done for this case with the result that both charges produce laser pulses of similar energy but differ in the structure, pulse length and average power.

Further investigations have been undertaken in order to verify the combined operation of two different bunch charges if the nominal design of the XFEL injector is extended by means of an additional laser system at the cathode. It was found that an additional laser system would enable a significant improvement of the parameters of both bunch charges in the same train. In particular it would allow the operation of the 250pC/500pC bunch pair at the working point which keeps the emittance growth (compared to the possible minimum) below 3% for both charges and the mismatch between the optical functions of the bunches below 1.1.  Other combined working points have been estimated for the bunch charges of 20pC, 100pC, 250pC, 500pC and 1nC. It was found that the operation with any bunch pair becomes possible if the nominal design of the XFEL injector is extended by an additional laser system.

A proper compression scenario is the next essential issue for a successful SASE generation. While the first rough choice of the RF settings can be done by means of the semi analytical model and rf tweak tool [7][6] the simulations of the full 6D dynamics require a more accurate correction of the rf settings later since different bunch charges experience different forces due to space charge, wake fields and CSR in BCs.

**Appendix:**

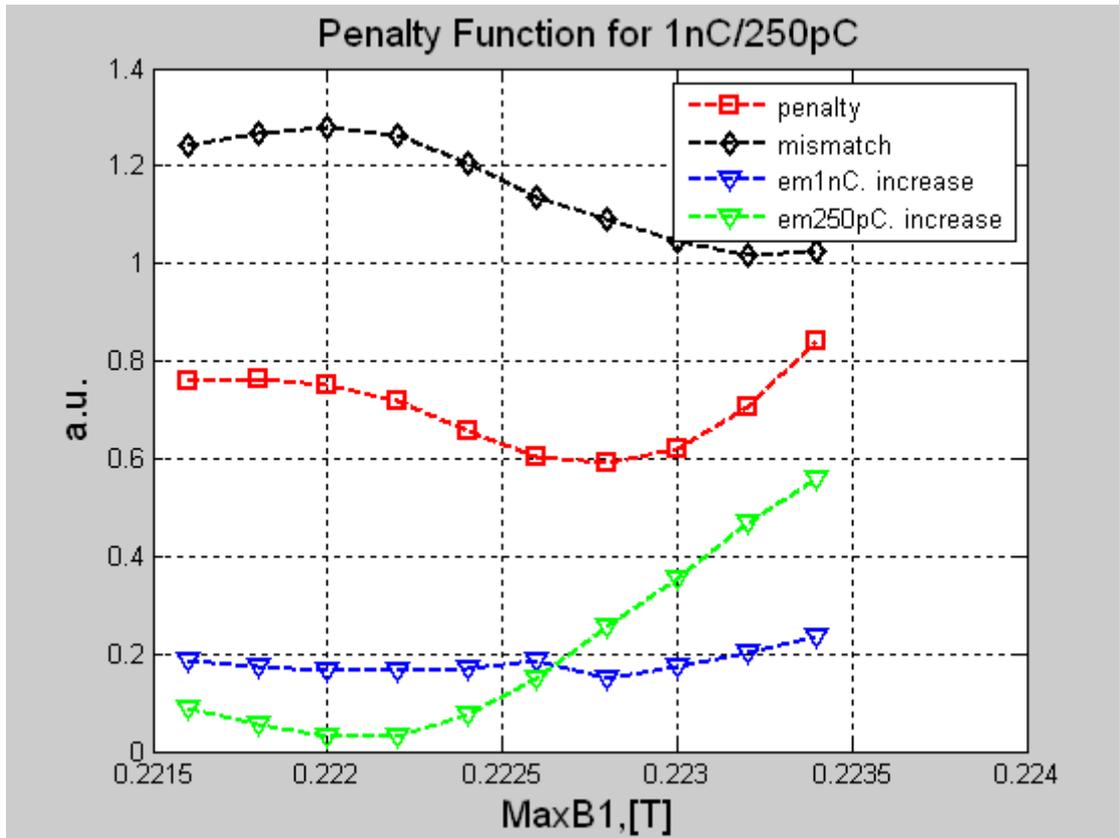

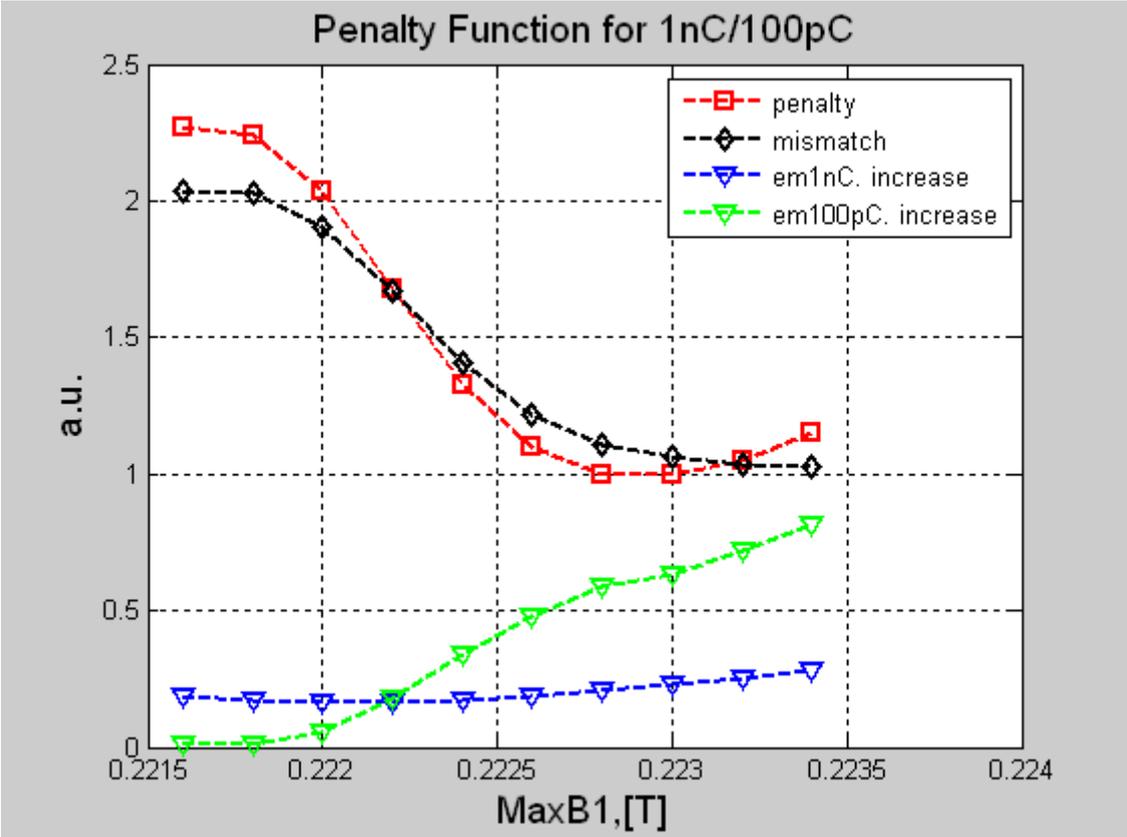

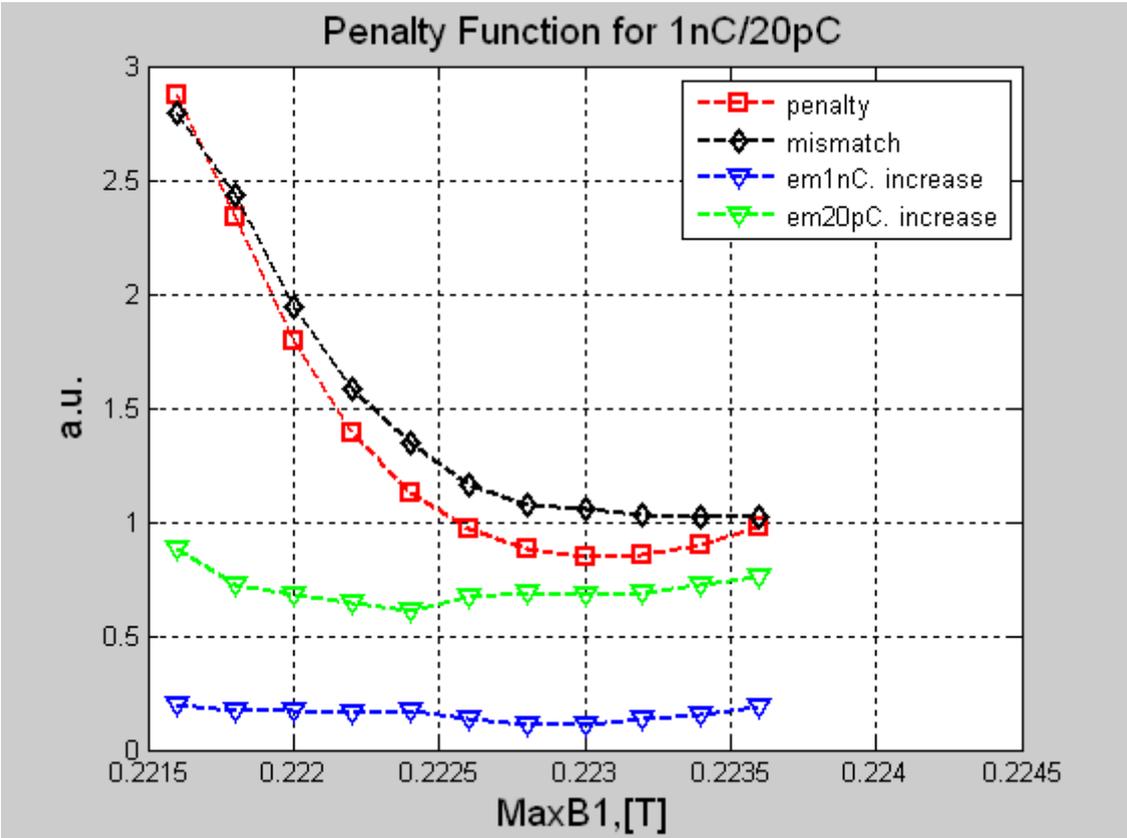

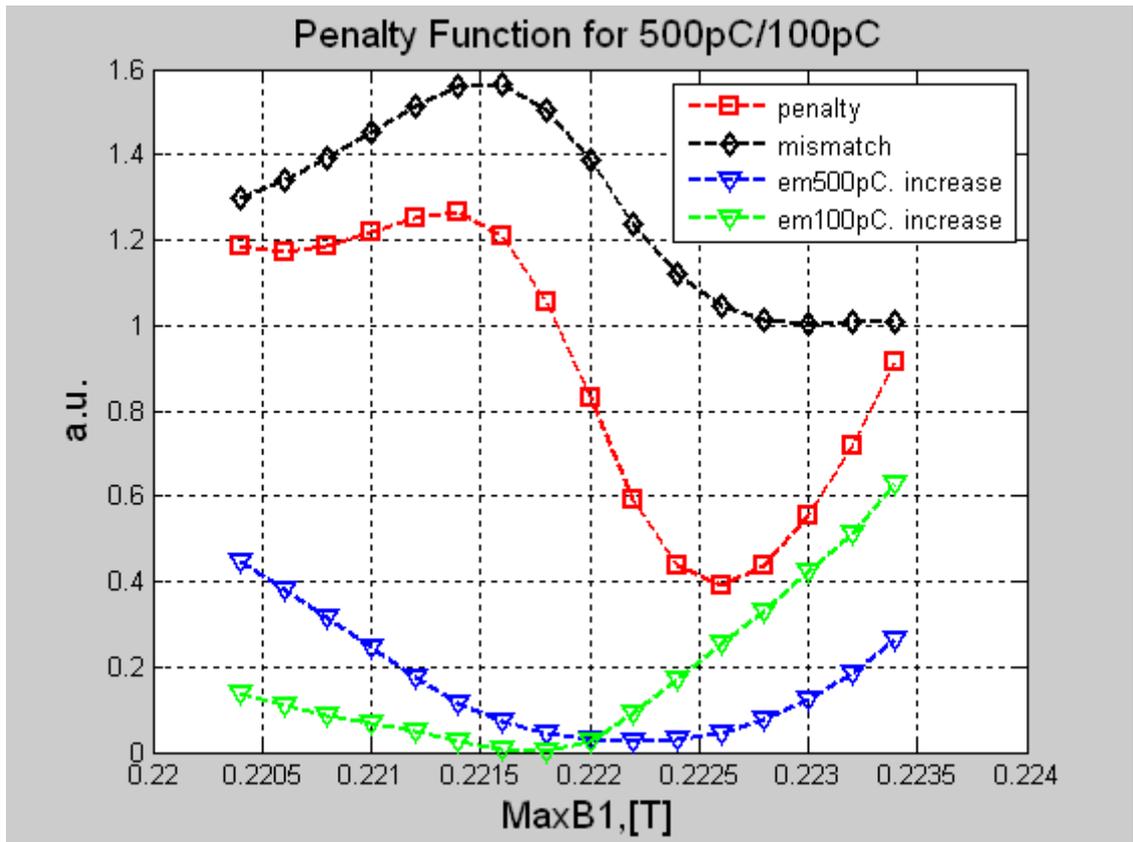
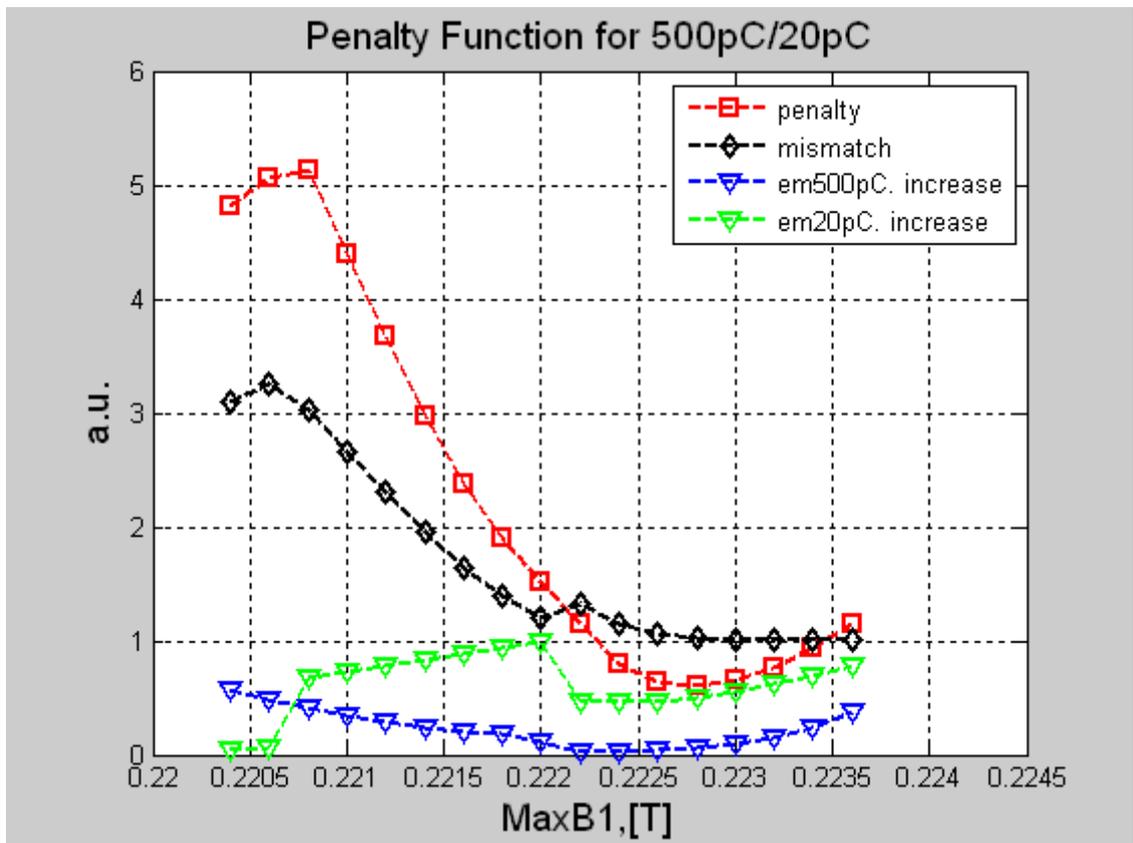

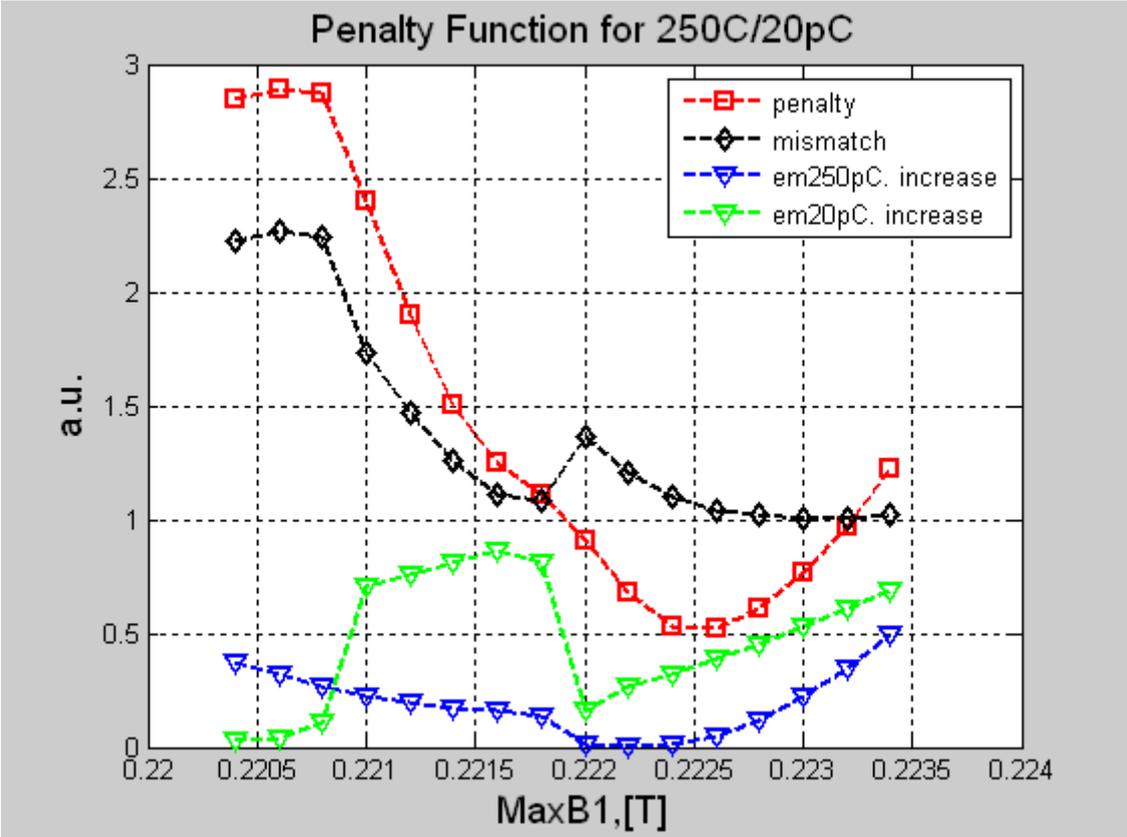

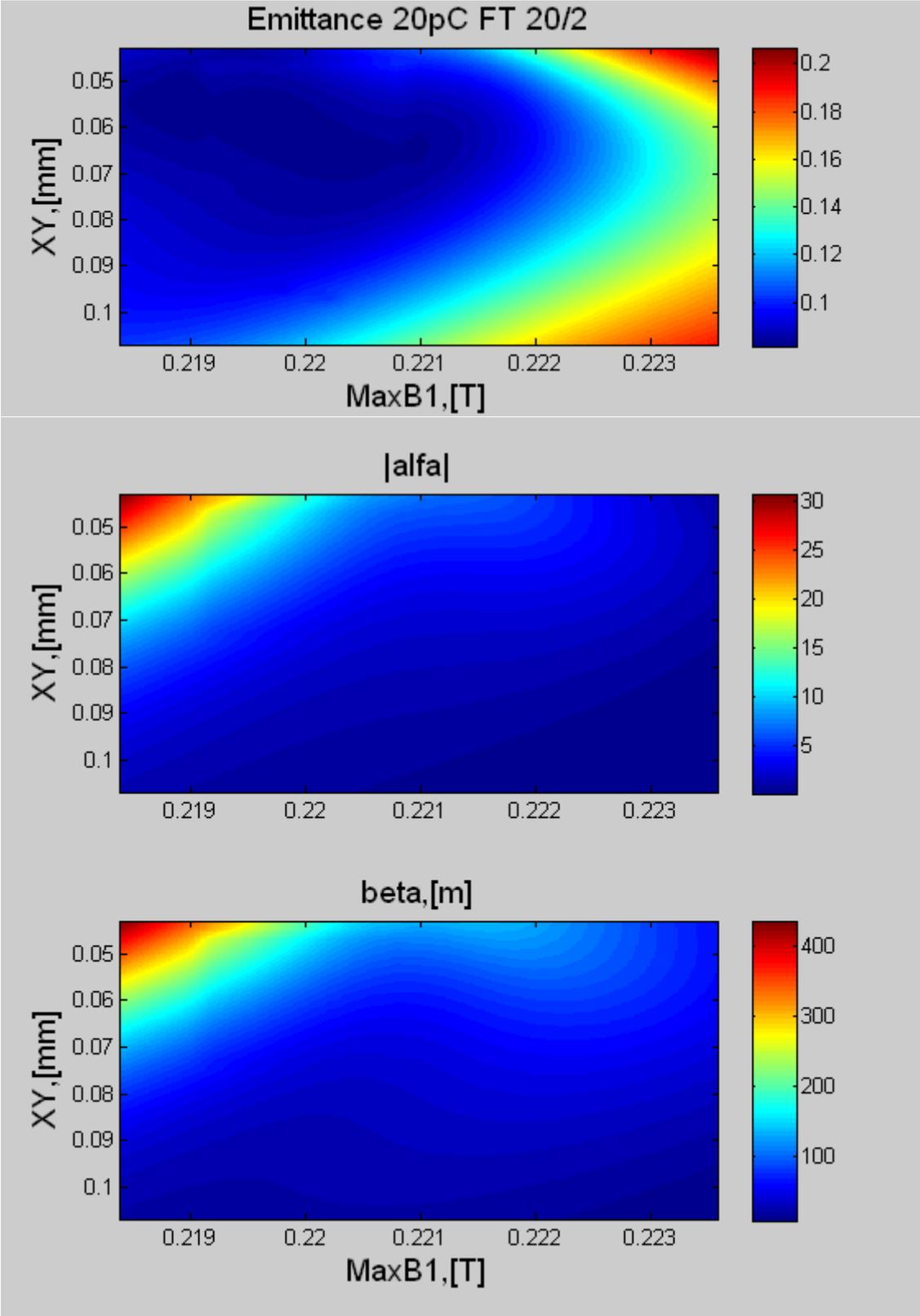

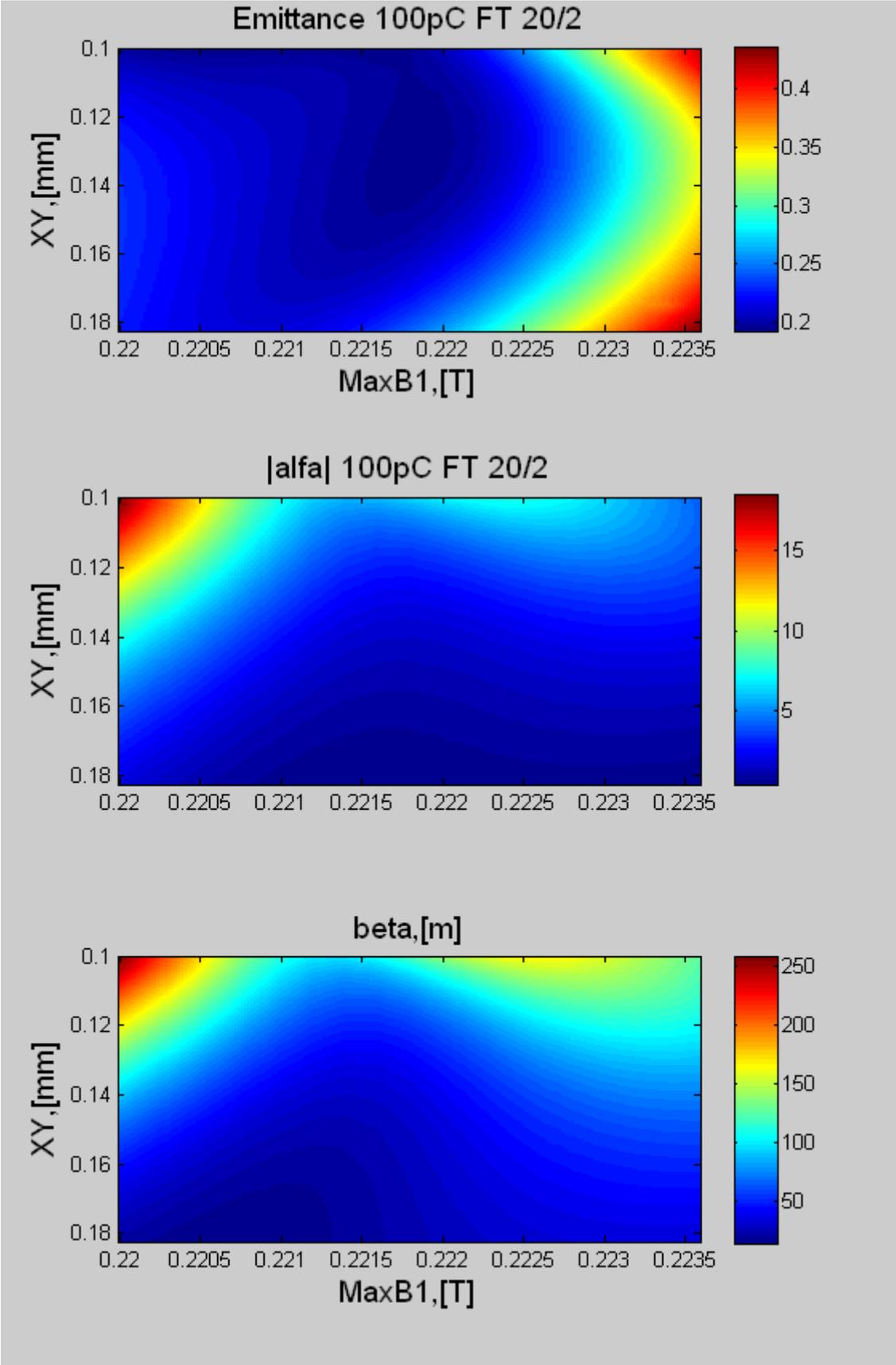

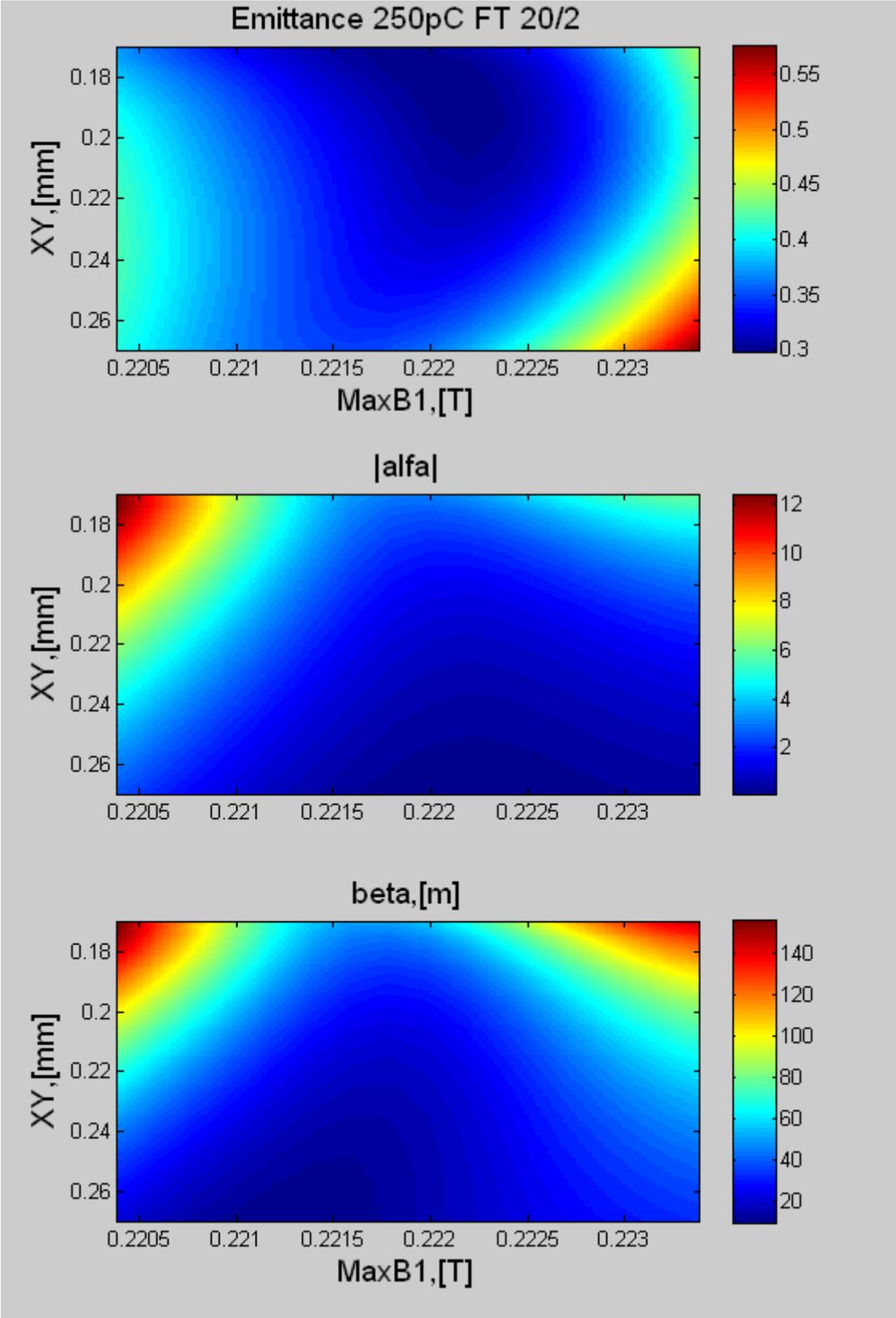

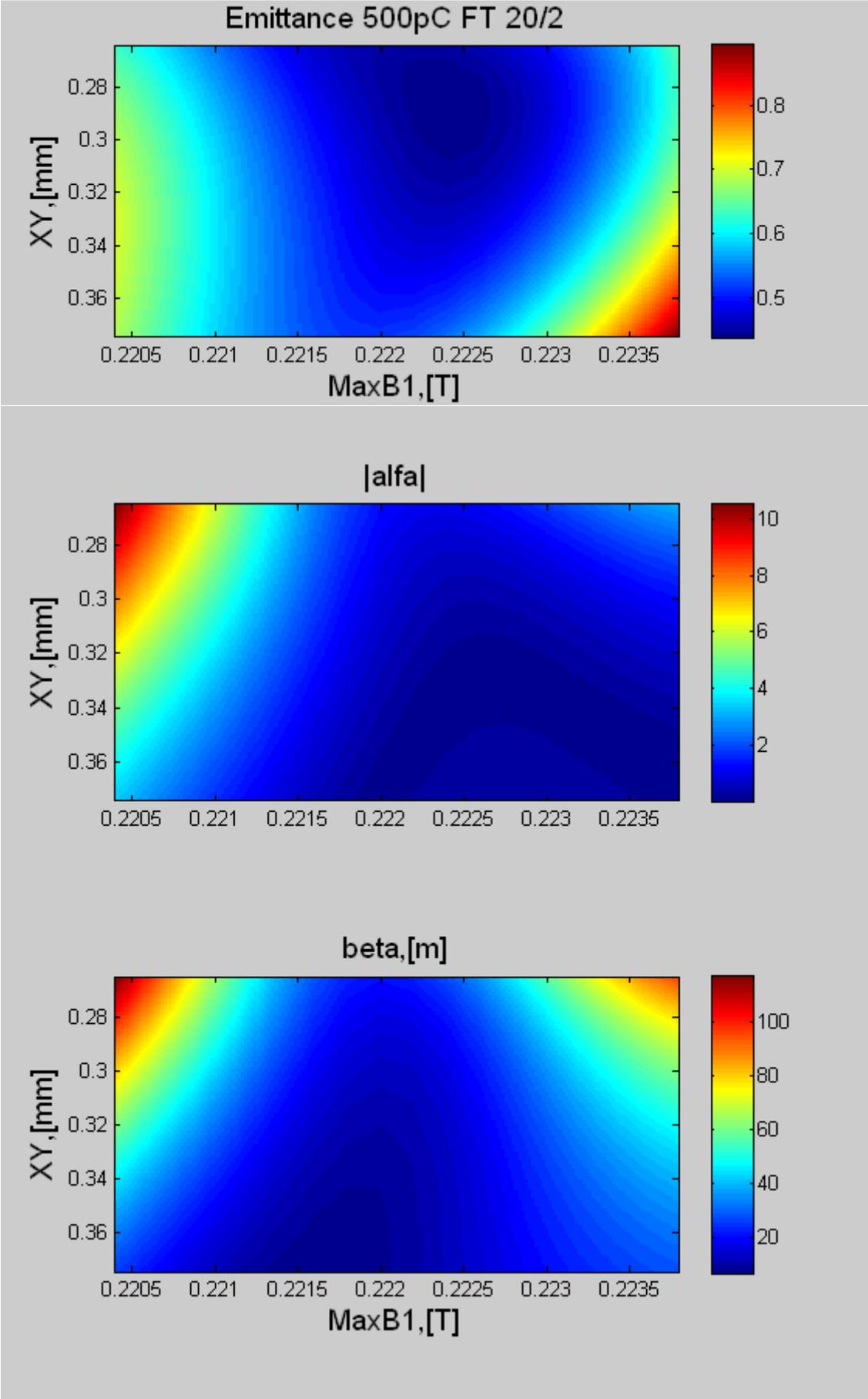

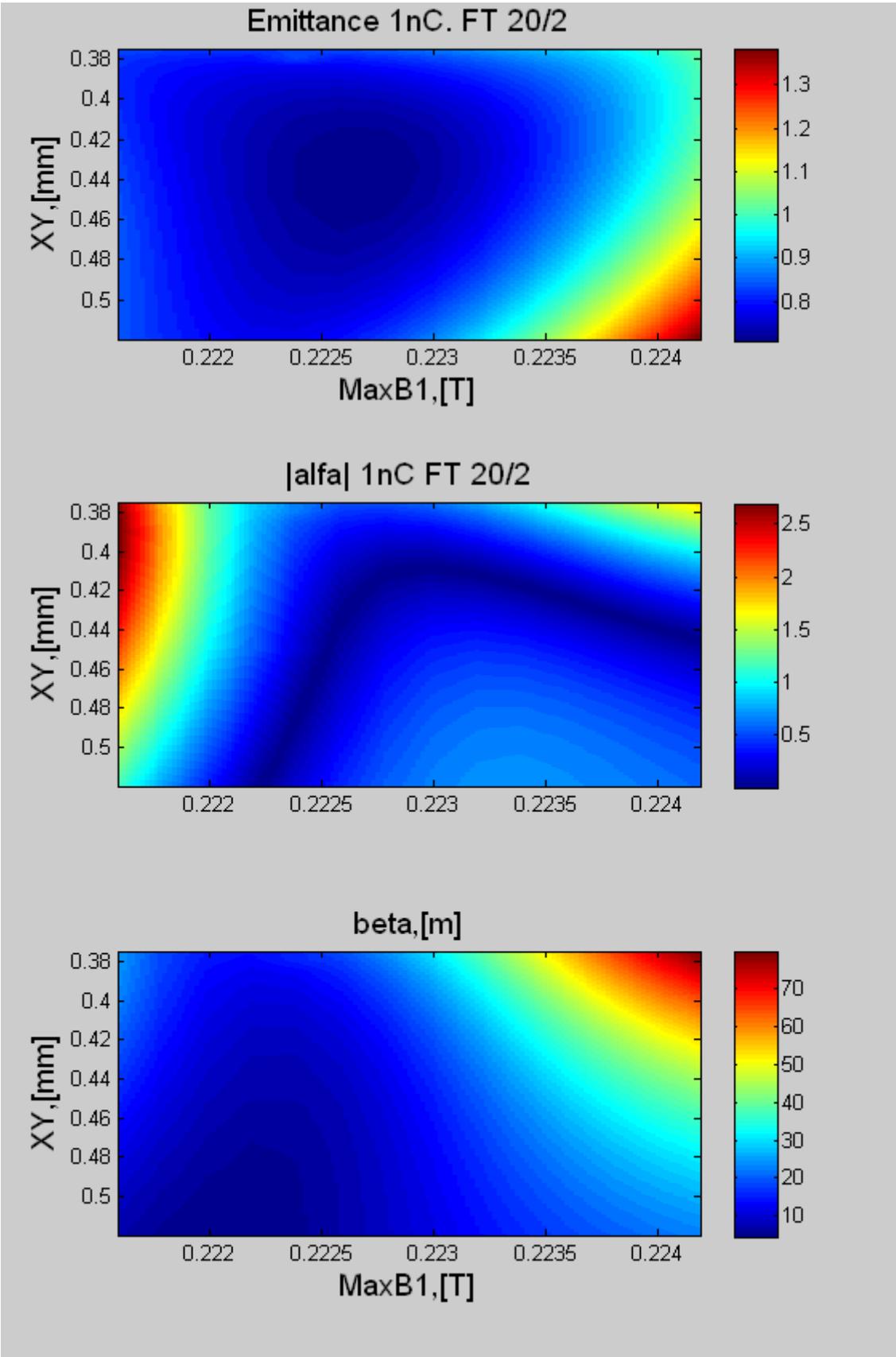